\documentclass[12pt]{iopart}

\usepackage{iopams}
\usepackage{graphicx}
\usepackage{caption}
\usepackage{subcaption}
\usepackage{hyperref}
\usepackage{hypernat}

\begin{document}
\title{Memory effects induce structure in social networks with activity-driven agents}
\author{A D Medus and C O Dorso}
\address{Departamento de F\'isica, Facultad de Ciencias Exactas y Naturales,
 Universidad de Buenos Aires, Pabell\'on 1, Ciudad Universitaria, Ciudad Aut\'onoma de Buenos Aires (1428), Argentina and IFIBA - CONICET}
\eads{admedus@df.uba.ar and codorso@df.uba.ar}

\begin{abstract}
Activity-driven modeling has been recently proposed as an alternative growth mechanism for time varying networks,
displaying power-law degree distribution in time-aggregated representation.
This approach assumes memoryless agents developing random connections, thus leading to random networks that fail
to reproduce two-nodes degree correlations and the high clustering coefficient widely observed in real social networks.
In this work we introduce these missing topological features by accounting for memory effects on 
the dynamic evolution of time-aggregated networks.
To this end, we propose an activity-driven network growth model including a triadic-closure step 
as main connectivity mechanism. 
We show that this simple mechanism provides some of the fundamental topological features expected for real social networks.
We derive analytical results and perform extensive numerical simulations in regimes with and without population growth.
Finally, we present two cases of study, one comprising face-to-face encounters in a closed gathering, 
while the other one from an online social friendship network.
\end{abstract}
\submitto{JSTAT}
\maketitle

\section{Introduction}\label{intro}
Social networks represent the different substrates on which we develop many aspects of our lives.
Knowledge, news, rumours and diseases are transmitted through an intricate social framework usually represented
by complex networks, which explains the growing interest of scientific community in such complex systems. 

Some topological characteristics of complex networks, as node degree distribution,
clustering coefficient, average shortest path length, modularity and degree-degree correlations have proved to be related 
with their spreading properties \cite{Pastor-Satorras01,KupermanAbramson,Newman1,Community,Miller2009,Volz11}.
It follows that a rigorous topological description of complex networks is a key matter in this area of knowledge. 
In particular, many kinds of human acquaintance networks display heterogeneous degree distribution (mainly power-law),
high clustering coefficient \cite{Newman2003}, strong communality \cite{GirNew02,Medus2005} and positive two-nodes degree correlation
(i.e. degree assortativity) \cite{Newman02} as distinctive features. 
These findings have promoted the emergence of different network growth mechanisms designed in order to provide accurate modeling tools.

Preferential attachment (\textbf{PA}) constitutes one of the most widespread heuristic mechanism giving rise to networks
with the ubiquitous power-law degree distribution (\textit{scale-free }networks) \cite{Barabasi99,Krapivsky.Redner2000}. 
It is an important part of the so called connectivity-driven network growth models and is empirically supported through
snapshots of particular collaboration or technological networks \cite{Barabasi99,Newman2001,Barabasi2002}.

Time-varying networks have been recently raised as a dynamic variant of the original static network representation. 
This new approach accounts for the changing nature of many empirical networks, comprising time-varying interactions by 
continuously creating and erasing edges between nodes.
In this context, an alternative mechanism based on the concept of \textit{activity rate} was recently proposed in order to explain 
the scale-free feature from a dynamic perspective \cite{Perra12}. 
The activity rate describes the degree of participation of a given individual in a particular social network. Participation can 
account for published papers in the case of scientific collaboration networks, movies and TV series filmed in actor networks, or 
messages shared in the Twitter microblogging network.
In each time step $\Delta t$ nodes are activated or not with probability proportional to its activity rate. Active nodes perform
random connections to another nodes chosen over the entire population. 
Then, while the greater the activity rate of an agent, the greater his accumulated acquaintances in a given time window.

The activity-driven model (\textbf{AD}) assumes memoryless agents that only perform uniformly random connections, 
properly reproducing the time dynamic of contacts and giving place to time-aggregated random scale-free networks \cite{Perra12}.
However, \textbf{AD} fails to replicate the high clustering coefficient 
and degree assortativity characteristics of many social networks, as shown in \cite{Starnini2013}.

Beyond their particular characteristics, every complex network is the result of an aggregation 
process over a given time window. As a consequence, network growth processes should be able to reproduce the main topological 
features observed in real social networks. 
The vast majority of empirical datasets suggest the need for some 
local connectivity mechanism that promotes transitive ties between social agents \cite{Watts06,Szell09,Szell10,Thurner13}.
To do so, the agents should remember their previous contacts as necessary condition.

In this work we introduce a generalized stochastic growth model (\textbf{GSG}) that provides some of the fundamental 
topological features expected for real social networks. In particular, those corresponding to higher-order correlations as
degree assortativity and the high average clustering coefficient.
\textbf{GSG} model assumes a population of agents with heterogeneous activity rates and long-term memory, i.e., 
they can remember their former acquaintances. This last characteristic aims to introduce the
impact of agents' current social environment on their further social development.
For this purpose, \textbf{GSG} model imposes a strictly local connectivity mechanism based on a combination of random ties together 
with a triadic-closure (\textbf{TC}) step that is well known for adding structure to the network
\cite{Barabasi2000,Davidsen02,Holme2002,Vazquez03,Saramaki06}.
Here we will focus on the evolution and topological features of time-aggregated networks under regimes with and without population growth.

The rest of this paper is organized as follows: In \sref{model} we introduce the details of the \textbf{GSG} model. In \sref{analytical}
we present the analytical treatment for the degree distribution. In Sections \ref{constant} and \ref{growth} we present exact and approximate
analytical solutions together with extensive numerical simulations for constant and growing population, respectively. 
Degree-degree correlations and clustering are studied by means of numerical simulations in \sref{higher}. 
Finally, in Section\ref{cases} we analyze two real social networks, the first corresponding to face-to-face encounters in a closed gathering 
under constant population and the second to a subgraph of Facebook online friendship network with population growth.
In \sref{discussion} all relevant results are summarized and discussed. 
\section{The model}\label{model}
Following the traditional network representation, nodes and edges correspond respectively to individuals and their ties in a social context.
Let $G_t(N;L)$ represents a network, or graph, composed of $N(t)$ nodes and $L(t)$ edges at time $t$. \textbf{GSG} assign to each node $i\in\{1,...,N\}$ an 
activity rate $a_i$ from a given activity pdf $F(a)$.
In the context of \textbf{GSG}, activity $a_i$ represents the rate at which new edges emerge from node $i$. 

Networks grow by addition of nodes and edges. 
Each added node is connected to another one randomly chosen from the current population. 
Additionally, the edges are introduced with heterogeneous rate given by:
\begin{equation}
 \beta(t)=\sum_{i=1}^{N(t)}a_i\:\:.
\end{equation}
Then, $\{L_t\}_{t\in\mathbb{R}_{\ge0}}$ defines a continuous time stochastic process with time-dependent rate $\beta(t)$. 
In order to simplify the subsequent analysis we define an embedded discrete stochastic process $\{L_{\ell}\}_{\ell\in\mathbb{N}}$
with $\ell$ the aggregated number of added edges. In this way, we also define the nodes' population growth rate $\gamma$
in terms of the edges' population growth rate (now formally equal to $1$). Hence, the evolution of the total number of edges and nodes  
are respectively given by $L(\ell)=L_0+(1+\gamma)\ell$, where $\gamma\ell$ comes from those edges associated with added nodes,
and $N(\ell)=N_0+\gamma\ell$, with $L_0$ and $N_0$ the initial values.

\subsection{Triadic-closure mechanism}
In \cite{Perra12} nodes are activated with a probability proportional to their activity rate.
From each active node $m$ edges arise, that will be connected to other nodes (actives or not) uniformly at random. 
This process gives rise to a time-aggregated random network with degree distribution inherited from the corresponding
activity density function $F(a)$.
Under this framework each node acts without memory of its previous connections, clearly at odds with almost all empirical evidence
(we analyze two examples in \sref{cases}).

Instead of multiple random connections, in \textbf{GSG} edges are added one by one following a mixed connection mechanism. 
The source node for each edge is selected proportionally to its activity rate, while the target node is chosen by 
the following procedure: a) with probability $q$ a second-neighbour of the source node is chosen in order 
to ``close a triangle'' (\textbf{TC} mechanism), or b) a random target node is selected with probability $(1-q)$.

Triadic-closure mechanism has been observed in many real social networks and is widely 
recognized as one of the most direct and natural ways to introduce transitivity (or clustering) in network growth models
\cite{Watts06,Davidsen02,Holme2002,Vazquez03,Saramaki06,Moriano13,Kleinberg2007}. 
It can also be understood as a way to replicate what we often experience in our social relations, namely, that 
usually new acquaintances are introduced to us through our current social environment. It is in this sense that we refer 
to long-term memory in agents, since everyone has an internal register of their previous contacts.
Agents' memory may also suggest the recurrence of previous contacts giving place to weighted edges. 
Nevertheless, here we are focused on addressing 
the network development under a parsimonious approach with unweighted edges (see \cite{Perra14}
for the relation of long-term memory with a reinforcement process).

\begin{figure}
        \centering
        \begin{subfigure}[b]{0.32\textwidth}
                \includegraphics[width=\textwidth]{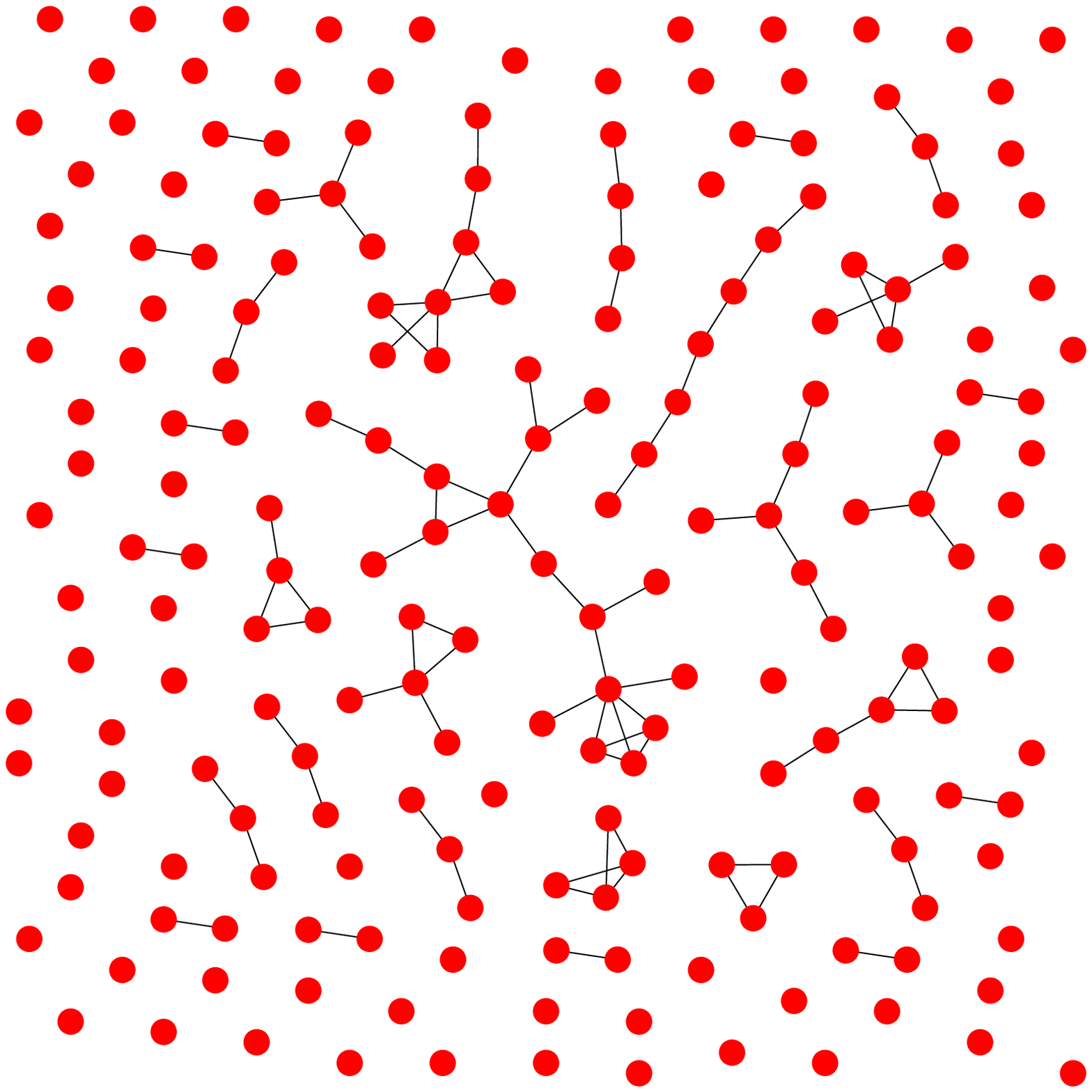}
                \caption{$\ell=100$}
                \label{fig:100}
        \end{subfigure}%
        ~ 
        \begin{subfigure}[b]{0.32\textwidth}
                \includegraphics[width=\textwidth]{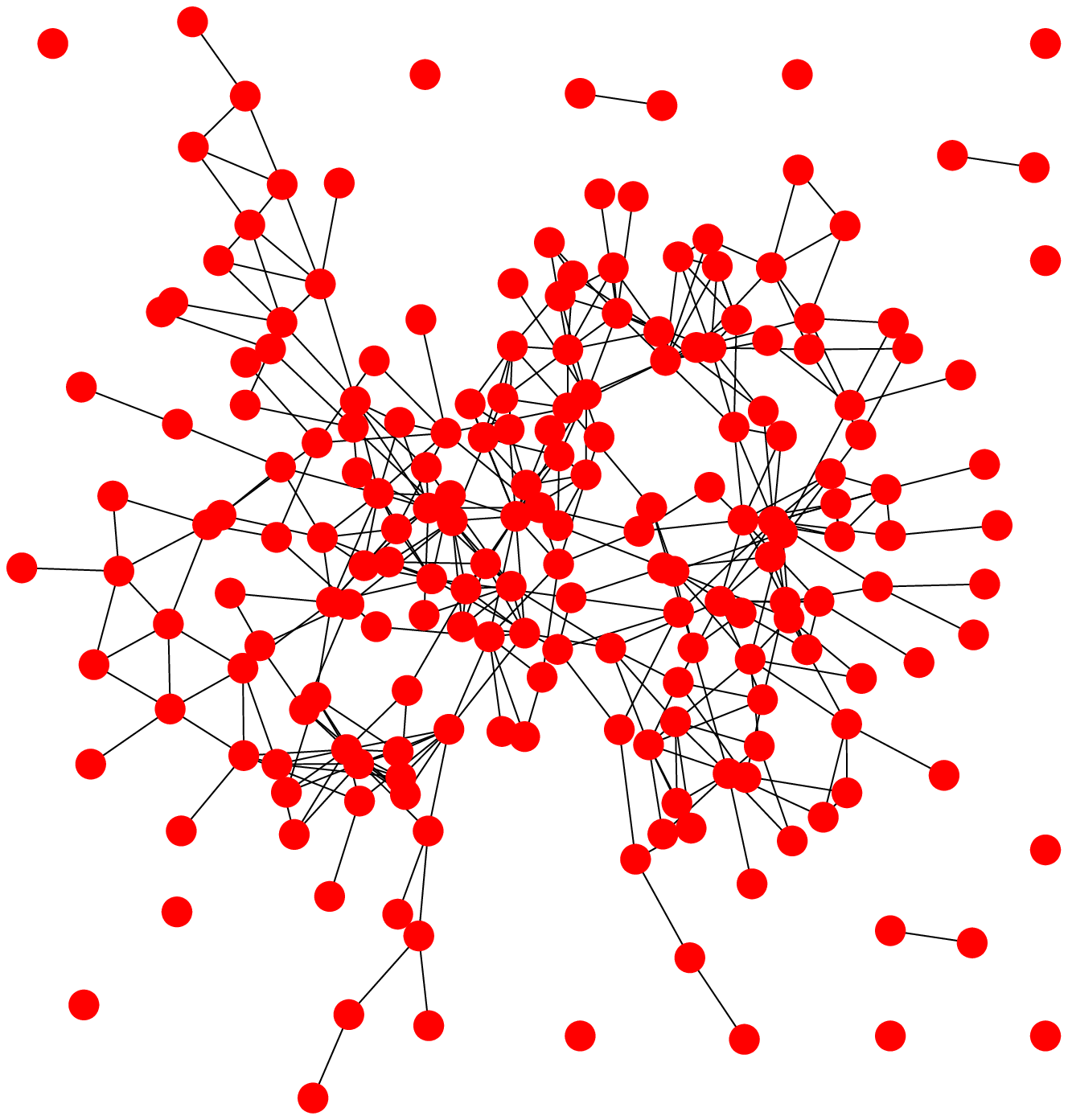}
                \caption{$\ell=400$}
                \label{fig:400}
        \end{subfigure}
        ~ 
        \begin{subfigure}[b]{0.32\textwidth}
                \includegraphics[width=\textwidth]{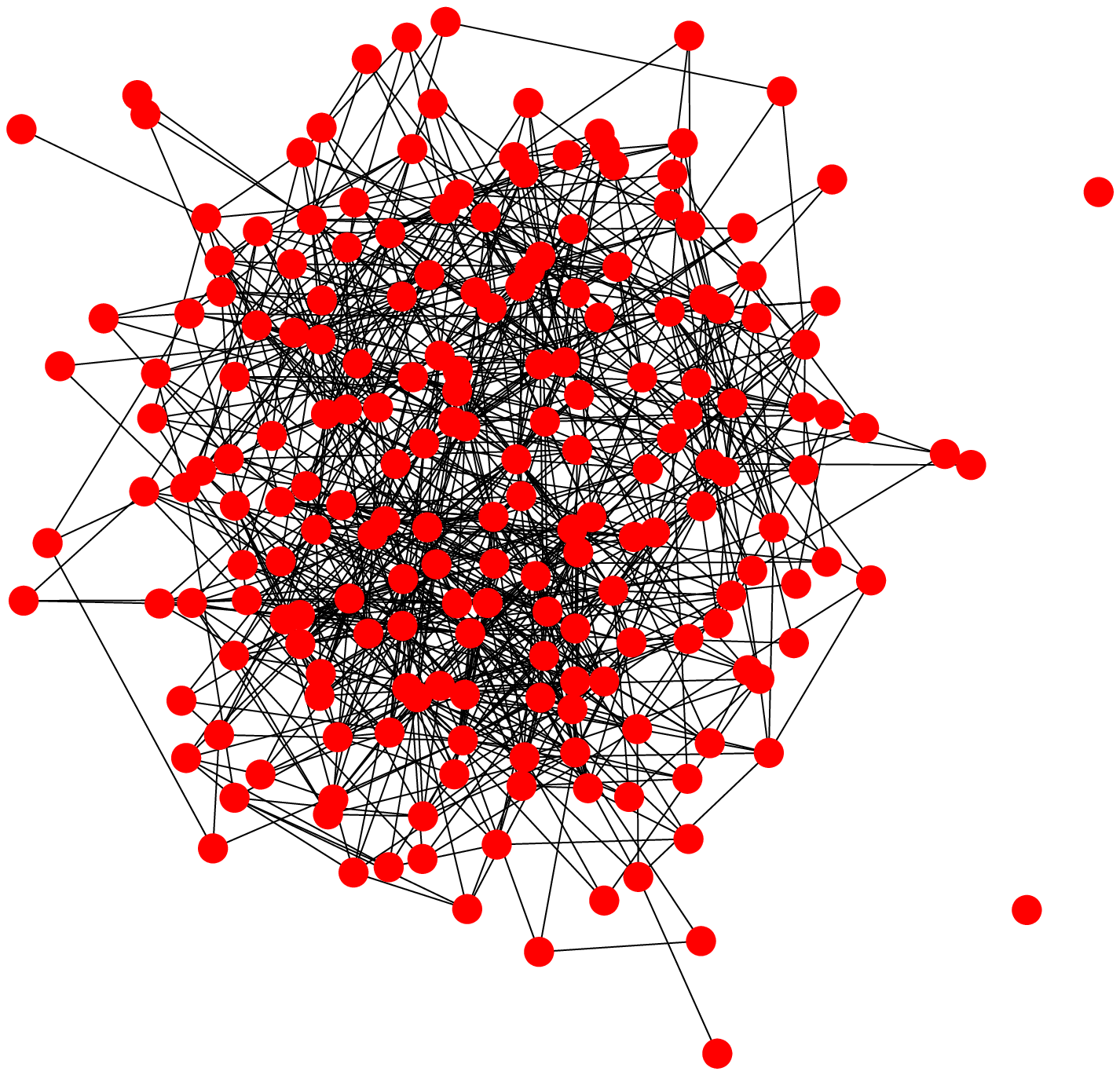}
                \caption{$\ell=1000$}
                \label{fig:1000}
        \end{subfigure}
        \caption{An example of three network growth stages (in terms of $\ell$) are shown for \textbf{GSG} model under 
        constant population with $N=200$ and triadic-closure probability $q=0.7$.}
        \label{fig0}
\end{figure}
Finally, the algorithmic description of \textbf{GSG} is summarized as follows:
\begin{itemize}
 \item [i. ] Choose a node at random with probability proportional to its activity and connect it as follows: a) with probability $q$, 
 to one of its second near-neighbours chosen uniformly at random,  (\textbf{TC} mechanism) or 
 b) with probability $(1-q)$, to another node
 chosen uniformly at random (random edge).
 No multiple connections are allowed. Repeat this step until one edge has been effectively connected.
 \item [ii. ] With probability $\gamma$ introduce a new node to the network and connect it with another one chosen uniformly at random.   
 \item [iii. ] Increment $\ell=\ell+1$ and repeat i) and ii) until the average degree reaches the desired value 
 $\langle k \rangle=2(L_0+(1+\gamma)\ell)/(N_0+\gamma \ell)$, being $N_0$ and $L_0$ the initial number of nodes and edges, respectively.
\end{itemize}
An example of the connectivity evolution for \textbf{GSG} model is shown in \fref{fig0}. The impact of \textbf{TC} mechanism
is clearly evidenced in this sequence.

In next sections we present a detailed analytical formulation of the model in order to obtain the degree 
distribution under two regimes: 
i) with constant population ($\gamma=0$) and ii) with population growth ($\gamma>0$).

\section{Analytical formulation}\label{analytical}
We begin defining the elements of \textbf{GSG} analytical formulation on a general framework. Let $P_{k,a}(\ell,\ell_0)$ be
the probability that a node introduced at step $\ell_0$ with activity rate $a$ has degree $k$ for a subsequent algorithm step 
$\ell>\ell_0$. Now we can define $\bar{N}_{k,a}(\ell)$ the mean number of nodes with degree $k$ and activity rate $a$ as
\begin{equation}
 \bar{N}_{k,a}(\ell)=\sum_{\ell_0=1}^{\ell}P_{k,a}(\ell,\ell_0)\:\:.
\end{equation}
The evolution of $\bar{N}_{k,a}(\ell)$ can be described from a continuum approach through a system of coupled rate equations
\cite{Krapivsky.Redner2000,KrapRedner01} as follows:
\begin{equation}\label{eq1}
\fl\eqalign{
\frac{\rmd\bar{N}_{k,a}}{\rmd \ell}=&q\left(\Theta(k-1,a,\ell)\bar{N}_{k-1,a}-\Theta(k,a,\ell)\bar{N}_{k,a}\right)+\frac{1-q}{N(\ell)}\left(\frac{a}{\langle a\rangle}+1\right)\left( \bar{N}_{k-1,a}-\bar{N}_{k,a}\right)\\
   &+\frac{\gamma}{N(\ell)}\left( \bar{N}_{k-1,a}-\bar{N}_{k,a}\right)+\gamma\delta_{k1}\:,
   }
\end{equation}
where the first term in the right-hand-side of \eref{eq1} is associated with the \textbf{TC} mechanism contribution to $\bar{N}_{k,a}(\ell)$ 
(with $\Theta(k,a,\ell)$ the \textbf{TC} kernel), while the second one corresponds to random edges contribution. This last term can be 
decomposed into the contribution of the source node chosen with probability proportional to its activity,
\begin{equation}
 \frac{1-q}{N(\ell)}\left(\frac{a}{\langle a\rangle}\right)\left( \bar{N}_{k-1,a}-\bar{N}_{k,a}\right)
\end{equation}
which is added to the contribution of the random connected target node, given by
\begin{equation}
 \frac{1-q}{N(\ell)}\left( \bar{N}_{k-1,a}-\bar{N}_{k,a}\right)\:.
\end{equation}
Continuing with our description of equation \eref{eq1}, the third term in its right-hand-side is the contribution of the remaining end of the edge added 
together with every $a$-activity new node (also tied uniformly at random). Finally, the last term $\gamma\delta_{k1}$ 
comes from the initial degree ($k=1$) of each $a$-activity new node itself. 
The \textbf{TC} kernel $\Theta(k,a,\ell)$ in \eref{eq1} is defined as:
\begin{equation}\label{eq2}
 \Theta(k,a,\ell)=\frac{a}{\langle a\rangle N(\ell)}+\frac{k}{2L(\ell)}\:.
\end{equation}
Every \textbf{TC} edge is tied to a source node chosen at random with probability proportional to its activity rate $a$,
leading to the first term in right-hand-side of \eref{eq2}, whereas the last one represents the preferential attachment term
 arising from \textbf{TC} target \cite{Holme2002,Vazquez03}.
After regrouping terms, equation \eref{eq1} can be rewritten as  
\begin{equation}\label{eq3}
  \frac{\rmd\bar{N}_{k,a}}{\rmd \ell}=\bar{N}_{k-1,a}\Phi_{\gamma,q}(k-1,a,\ell)-\bar{N}_{k,a}\Phi_{\gamma,q}(k,a,\ell)+ \gamma\delta_{k1}
\end{equation} 
being $\Phi_{\gamma,q}(k,a,\ell)$ the generalized connectivity kernel given by
\begin{equation}\label{eq3*}
 \Phi_{\gamma,q}(k,a,\ell)=\frac{1}{N(\ell)}\left(\frac{a}{\langle a\rangle}+1-q+\gamma\right)+\frac{k}{2L(\ell)}q\:.
\end{equation}
 
Finally, the resulting  expression for the mean population of nodes with degree $k$ ($\bar{N}_k(\ell)$) comprising all 
possible activity rates with pdf $F(a)$ is given by
\begin{equation}\label{eq4}
 \bar{N}_k(\ell)=\int_\Omega F(a)\bar{N}_{k,a}(\ell) \rmd a
\end{equation}
being $\Omega$ the domain of $F(a)$. Now we can formally define the degree distribution function for 
a given quantity of aggregated edges $\ell$ as $P(k,\ell)=\bar{N}_k(\ell)/N(\ell)$.

\textbf{GSG} allows a broad flexibility in both activity distribution and population growth regimes. 
In next sections we will solve \eref{eq3} under constant population ($\gamma=0$) and population growth ($\gamma>0$) 
regimes, in order to bring out a detailed analysis of $P(k,\ell)$ in this cases. On the other hand, we will focus on two paradigmatic
cases for activity pdf: 
i) constant activity ($F(a)=\delta(a-a_0)$) as the most frequent assumption, and ii) power-law activity pdf ($F(a)\propto a^{-\xi}$)
recently found in some real social networks \cite{Makse13}.

\section{Constant population}\label{constant}

In the particular case of constant population, network growth takes place only through the addition of new edges between existing nodes. 
Thus, the coupled system of ordinary differential equations governing the evolution of $\bar{N}_{k,a}(\ell)$ can be obtained 
by substituting $\gamma=0$ in \eref{eq3} that, after replacing $\Phi_{0,q}(k,a,\ell)$, reads
\begin{equation}\label{eq5}
 \eqalign{\frac{\rmd\bar{N}_{k,a}}{\rmd \ell}=&\bar{N}_{k-1,a}\left[\frac{1}{N_0}\left(\frac{a}{\langle a\rangle}+1-q\right)+\frac{(k-1)q}{2L(\ell)}\right]-\\
                                              &\bar{N}_{k,a}\left[\frac{1}{N_0}\left(\frac{a}{\langle a\rangle}+1-q\right)+\frac{kq}{2L(\ell)}\right]}
\end{equation} 
where $N(\ell)=N_0$ $\forall \ell\geq0$ and $L(\ell)=L_0+\ell$. 
There is a natural constraint imposed to $k$ in \eref{eq5}, that is $k\leq(N_0-1)$. Moreover, the asymptotic solution to \eref{eq5}
adopt the trivial form $\bar{N}_{k,a}=N_0\delta_{k,(N_0-1)}$ $\forall a\geq 0$. However, here we are interested only 
in the non-trivial transient solution.

Equation \eref{eq5} can be solved in general by means of an iterative scheme as follows
\begin{equation}\label{eq6}
 \bar{N}_{k,a}(\ell)=\left(\bar{N}_{k,a}(0)+\int_0^\ell \frac{\bar{N}_{k-1,a}(\ell')}{\Pi_{k,a}(\ell')}\Phi_{0,q}(k-1,a,\ell')\:\rm{d}\ell'\right)\Pi_{k,a}(\ell)
\end{equation}
where $\Pi_{k,a}(\ell)=A\left(L_0+\ell\right)^{-kq/2}\exp\left(-\left(a/\langle a\rangle+1-q\right)\ell /N_0\right)$ is solution of
\begin{equation}\label{eq7}
\frac{\rmd\Pi_{k,a}(\ell)}{\rmd\ell}=-\Pi_{k,a}(\ell)\Phi_{0,q}(k,a,\ell).
\end{equation}

Closed form solutions to \eref{eq5} are only reached in some particular cases. 
For instance, it can be easily shown that if $q=0$, the solution to \eref{eq5} is 
\begin{equation}\label{eq5*}
 \bar{N}_{k,a}(\ell)=N_0\times \mathrm{Pois}\left(k;\lambda=\ell(a+\langle a\rangle)/(\langle a\rangle N_0)\right),
\end{equation}
where $\mathrm{Pois}(k;\lambda)=(\lambda^{k}/k!)\exp(-\lambda)$ is the Poisson distribution with mean $\lambda$. 
Nevertheless, we will acquire some insight about exact behaviour of $\bar{N}_{k,a}$ by analyzing approximate solutions 
to \eref{eq5} under extreme conditions.
To this end, we assume the condition 
\begin{equation}\label{conda}
\textbf{a.}\hspace{0.5cm} \frac{1}{N_0}\left(\frac{a}{\langle a\rangle}+1-q\right)\gg q\frac{k}{2(L_0+\ell)}\:.
\end{equation} 
By virtue of condition $\textbf{a.}$, the approximate solution to \eref{eq5} results
\begin{equation}\label{eq8}
 \bar{N}_{k,a}(\ell)\sim \frac{1}{k!}\left(\frac{a/\langle a\rangle+1-q}{N_0}\:\ell\right)^{k}\rme^{-\frac{a/\langle a\rangle+1-q}{N_0}\:\ell}
\end{equation}
i.e., a Poisson distribution with mean $\lambda=\ell(a/\langle a\rangle+1-q)/N_0$.

On the other hand, the opposite condition to (\ref{conda}) corresponds to assume
\begin{equation}\label{condb}
 \textbf{b.}\hspace{0.5cm} \frac{1}{N_0}\left(\frac{a}{\langle a\rangle}+1-q\right)\ll q\frac{k}{2(L_0+\ell)}\:,
\end{equation}   
yielding another approximate extreme solution to \eref{eq5} satisfying
\begin{equation}\label{eq8*}
 \fl \bar{N}_{k,a}(\ell)\sim\left(L_0+\ell\right)^{-q/2}\left[1-\left(\frac{L_0+\ell}{L_0}\right)^{-q/2}\right]^{k-1}\approx \left(L_0+\ell\right)^{-q/2}
  \rme^{-(k-1)\left(\frac{L_0+\ell}{L_0}\right)^{-q/2}}\:,
\end{equation}
showing a clear exponential decay with independence of the activity rate $a$. 
We shall see that another activity-independent solution is obtained again for large-$k$ values under population growth regime.
\begin{figure}
 \centering
 \includegraphics[width=15 cm,height=10 cm,keepaspectratio=true]{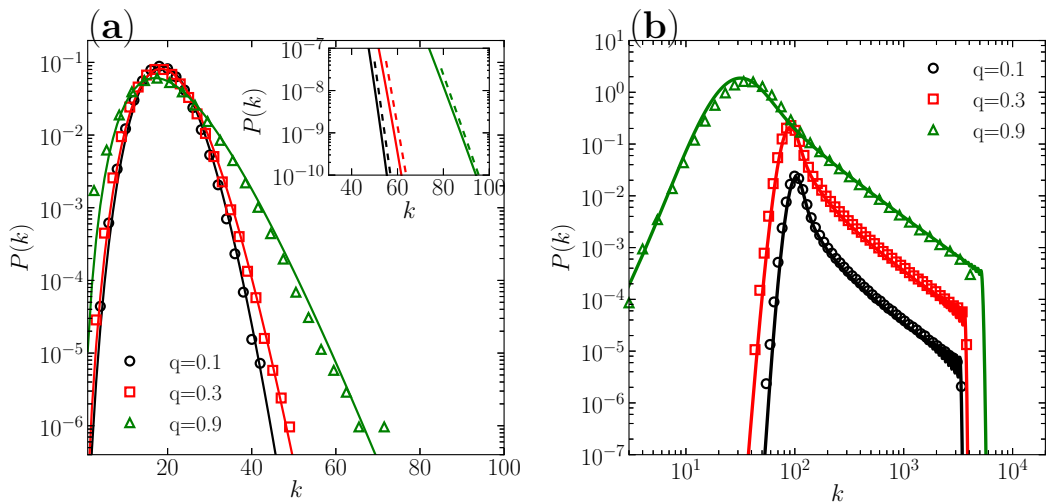}
 \caption{Degree distribution function $P(k)$ under constant population regime. \textbf{(a)} Activity pdf $F(a)=\delta(a-a_0)$ on networks with $N=10^5$ and $\langle k \rangle=20$, and
 \textbf{(b)} Activity pdf $F(a)\propto a^{-1.5}$ on networks with $N=10^5$ and $\langle k \rangle=200$.
 In both cases symbols correspond to averages over $100$ numerical simulations for: $q=0.1$ (black circles), $q=0.3$ (red squares) and $q=0.9$ 
 (green triangles). For the sake of clarity, the plots for different $q$ values have been shifted in all cases.
 Solid lines correspond to numerical solutions to \eref{eq6} subsequently integrated in \eref{eq4}, in order
to obtain $P(k,\ell)=\bar{N}_k(\ell)/N(\ell)$.
 \textbf{(Inset in (a))} Enlarged detail of high-$k$ values behaviour in order to compare the exponential decay constant for exact solutions (solid lines) with those corresponding to the 
 approximate formulation of \eref{eq8*} (dashed lines).}
 \label{fig1}
\end{figure}
Finally, $\bar{N}_k(\ell)$ is obtained by performing the integral of \eref{eq4} between $\bar{N}_{k,a}(\ell)$ 
for constant population and the activity pdf $F(a)$. 

We perform numerical simulations of \textbf{GSG} for constant population with $N=10^5$ nodes,
together with numerical integration of equations \eref{eq5} and \eref{eq4} in order to obtain the corresponding degree distribution 
$P(k,\ell)=\bar{N}_k(\ell)/N(\ell)$. We analyze two particular activity regimes: i) constant activity $F(a)=\delta(a-a_0)$, 
and ii) power-law activity pdf $F(a)\propto a^{-1.5}$. We show the very good agreement between simulations an theoretical 
predictions for constant population in \fref{fig1}. For constant and homogeneous activity, \fref{fig1}-a shows the expected
behaviour for $P(k)$, i.e., Poisson-like for small-$k$ with a marked exponential decay when condition (\ref{condb}) is satisfied 
(see inset in \fref{fig1}-a). Under power-law activity pdf, $P(k)$ is dominated by $F(a)$ power-law decay for mid-range $k$-values as shown 
in \fref{fig1}-b, while the limit cases are similar to those of the previous scenario.

\section{Population growth}\label{growth}
Now we study the population growth regime considering $\gamma>0$ in equation \eref{eq3}. This case is very relevant because 
growth constitute one of the fundamental assumptions to obtain scale-free networks from preferential attachment mechanism.
We will shown here that this feature is also present for \textbf{GSG} mechanism under population growth.

\begin{figure}
 \centering
 \includegraphics[width=15 cm,height=10 cm,keepaspectratio=true]{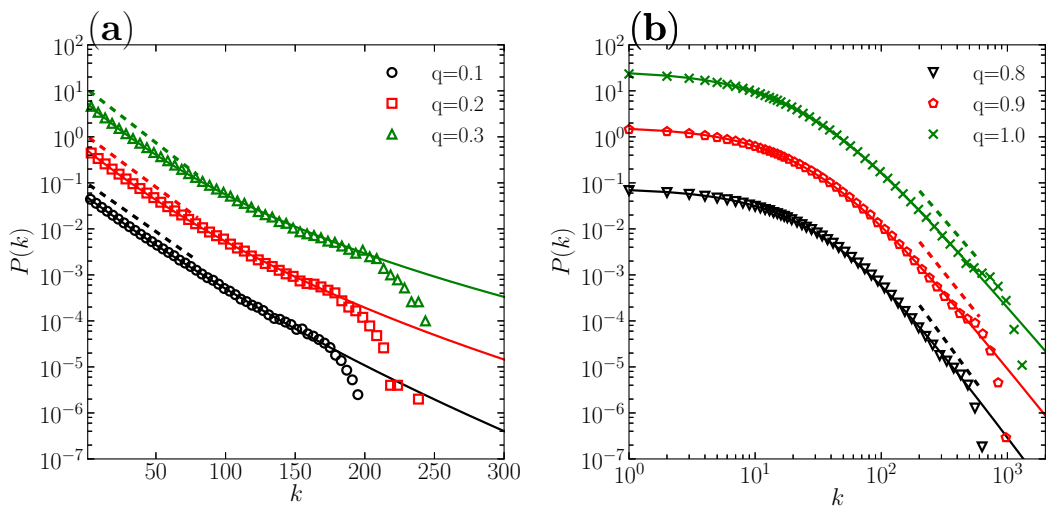}
 \caption{Degree distribution function $P(k)$ under population growth regime and $F(a)=\delta(a-a_0)$ for networks with 
 final population $N=10^5$ and $\langle k \rangle=20$. 
 Symbols correspond to averages over $100$ numerical simulations for: \textbf{(a)} $q\in\{0.1,0.2,0.3\}$, and \textbf{(b)} $q\in\{0.8,0.9,1.0\}$. 
 Solid lines correspond to numerical solutions of \eref{eq9} properly normalized in order to obtain $P(k)$.
 Asymptotic extreme solutions of \eref{eq11*} and \eref{eq11} are plotted in shifted dashed lines.
 For the sake of clarity, the plots for different $q$ values have been shifted.}
 \label{fig2}
\end{figure}

Regrouping terms in \eref{eq3} we can rewrite it now for $\gamma>0$ as
\begin{equation}\label{eq9}
 \eqalign{\frac{\rmd\bar{N}_{k,a}}{\rmd \ell}=&\bar{N}_{k-1,a}\left[\frac{1}{N(\ell)}\left(\frac{a}{\langle a\rangle}+1-q+\gamma\right)+\frac{(k-1)q}{2L(\ell)}\right]\\
 &-\bar{N}_{k,a}\left[\frac{1}{N(\ell)}\left(\frac{a}{\langle a\rangle}+1-q+\gamma\right)+\frac{kq}{2L(\ell)}\right]+\gamma\delta_{k1}}
\end{equation} 
where now $N(\ell)=N_0+\gamma\ell$ and $L(\ell)=L_0+(1+\gamma)\ell$.
Unlike the previous case, now is possible to obtain non-trivial asymptotic solutions to \eref{eq9}. 
This kind of solutions have the general form $\bar{N}_{k,a}(\ell)=n_{k,a} \ell$, with $n_{k,a}$ 
an unknown function of degree $k$ and activity rate $a$ \cite{Krapivsky.Redner2000}. 
Then, solving equation \eref{eq9} for $n_{k,a}$ under asymptotic condition ($\ell\rightarrow\infty$) we obtain
\begin{equation}\label{eq10}
 n_{k,a}=n_{1,a}\prod_{\mathtt{j}=1}^{k-1}\frac{2(\gamma+1)(a/\langle a\rangle+1-q+\gamma)+q\gamma \:\mathtt{j}}
 {2(\gamma+1)\left(a/\langle a\rangle+1-q+2\gamma\right)+q\gamma\: (\mathtt{j}+1)}
\end{equation}
where $n_{1,a}$ is the solution for $k=1$ given by
\begin{equation}\label{eq10*}
 n_{1,a}=\frac{2\gamma^2(\gamma+1)}{2(\gamma+1)\left(a/\langle a\rangle+1-q+2\gamma\right)+q\gamma}\:.
\end{equation}
Beyond its rigorous expression, $n_{k,a}$ takes very simple forms under approximate scenarios. 
The idea behind these approximations is to obtain a simplified framework where solutions have a more 
evident meaning than under its exact form.

\begin{figure}
 \centering
 \includegraphics[width=15 cm,height=10 cm,keepaspectratio=true]{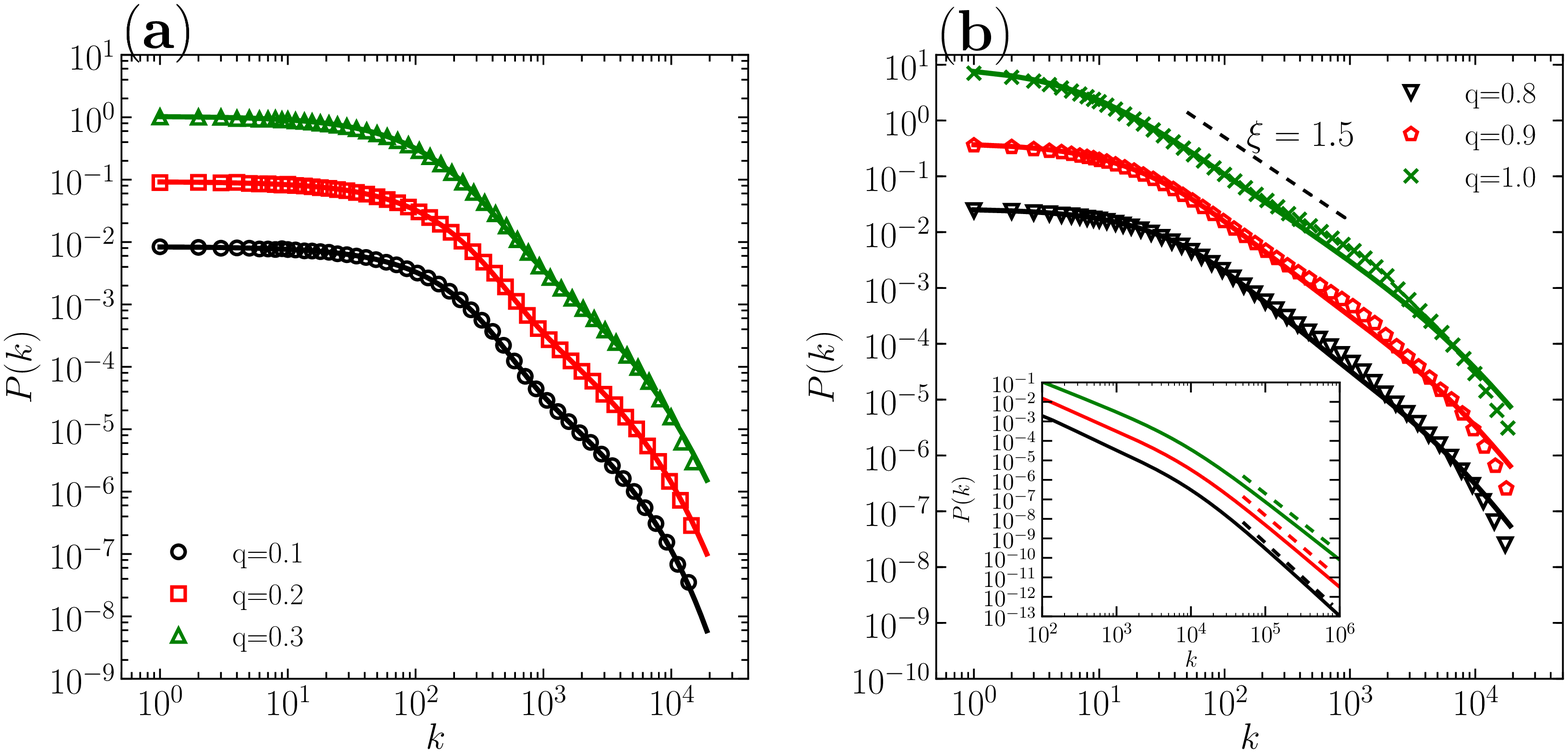}
 \caption{Degree distribution function $P(k)$ under population growth regime and $F(a)\propto a^{-1.5}$ 
 for networks with final population $N=10^5$ and $\langle k \rangle=200$. 
 Symbols correspond to averages over $100$ numerical simulations for: \textbf{(a)} $q\in\{0.1,0.2,0.3\}$, and \textbf{(b)} $q\in\{0.8,0.9,1.0\}$. 
 Solid lines correspond to numerical solutions of \eref{eq9} averaged through \eref{eq4}
 and properly normalized in order to obtain $P(k)$.
 Asymptotic extreme solutions of \eref{eq11*} and \eref{eq11} are plotted in shifted dashed lines.
 \textbf{(Inset in (b))} Enlarged detail of higher $k$ behaviour in order to compare the power-law decay for exact solutions (solid lines) with those corresponding to the 
 approximate formulation of \eref{eq12} (dashed lines).
 For the sake of clarity, the plots for different $q$ values have been shifted.}
 \label{fig3}
\end{figure}

The first approximate scenario corresponds to neglect preferential attachment terms in \eref{eq9}, assuming that
\begin{equation}\label{conda*}
 \textbf{a.}\hspace{0.5cm} \left(\frac{a}{\langle a\rangle}+1-q+\gamma\right)\langle k\rangle_\ell\gg qk
\end{equation}
being $\langle k\rangle_\ell=2L(\ell)/N(\ell)$ the mean degree after $\ell$ aggregated edges. 

Introducing approximation \textbf{a.} into \eref{eq9} and substituting again $\bar{N}_{k,a}(\ell)=n_{k,a} \ell$
we obtain the asymptotic solution under this approximate framework, 
\begin{equation}\label{eq11*}
 n_{k,a}\sim\left(1+\frac{\gamma}{a/\langle a\rangle+1-q+\gamma}\right)^{-(k-1)}
\end{equation}
which shows a pure exponential decay. 
Let us now analyze the alternative extreme condition
\begin{equation}\label{condb*}
 \textbf{b.}\hspace{0.5cm} \left(\frac{a}{\langle a\rangle}+1-q+\gamma\right)\langle k\rangle_\ell\ll qk\:.
\end{equation}
This last condition is satisfied when $k$ and $q$ are large enough, thus \eref{eq10} takes the form  
\begin{equation}\label{eq11}
 n_{k,a}\sim k^{-1-\frac{2(\gamma+1)}{q}}.
\end{equation}
showing that $n_{k,a}$ has power-law behaviour with exponent 
$\alpha=1+\frac{2(\gamma+1)}{q}$, resulting $\alpha>3$ when $0< q\leq 1$ and $\gamma > 0$.  
An important fact is that $n_{k,a}$ becomes absolutely independent of $a$. Accordingly, $\bar{N}_k(\ell)$ from
 \eref{eq4} adopt the same asymptotic power-law behaviour of $n_{k,a}$,
\begin{equation}\label{eq12}
 \bar{N}_k\sim k^{-1-\frac{2(\gamma+1)}{q}}.
\end{equation}
As a consequence of \eref{eq12}, the asymptotic large-$k$ behaviour of $\bar{N}_k$ is sustained only on \textbf{TC}
mechanism with total disregard of any particular activity distribution. Once again, the combination 
of population growth with a preferential attachment term (in this case coming from \textbf{TC} mechanism) gives place to power-law behaviour, 
just as in the original \textbf{PA} model \cite{Barabasi99}.
Unfortunately, this kind of behaviour seems to be hard to detect in real social networks because condition (\ref{condb*}) 
is only fulfilled for very large values ​​of $k$, which are rarely achieved or are perturbed by finite size effects.

We perform again extensive numerical simulations for \textbf{GSG} model but this time with population growth. In
\fref{fig2} we show the good agreement between the results of numerical simulations and theoretical predictions from \eref{eq10}
for the particular case of constant activity rate. Extreme solutions corresponding to both conditions analyzed in the text are 
represented in figures {\ref{fig2}}-a and {\ref{fig2}}-b by shifted dashed lines. A very good agreement between theory and simulations
is shown again in \fref{fig3}, now for power-law $F(a)$.

\section{Degree-degree correlation and clustering}\label{higher}
\begin{figure}
 \centering
 \includegraphics[width=15 cm,height=13 cm,keepaspectratio=true]{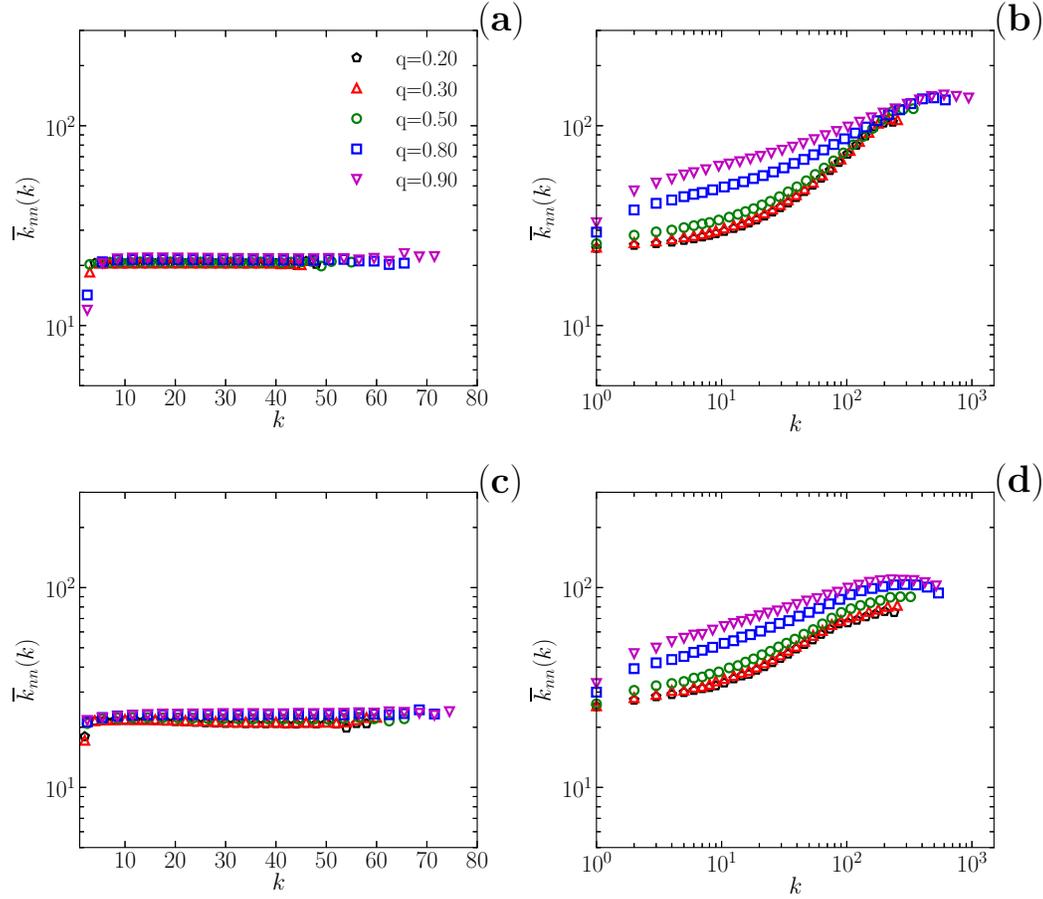}
 \caption{ Neighbours average degree $\bar{k}_{nn}(k)$ for
 networks obtained by \textbf{GSG} model, in all cases with final population $N=10^5$ and $\langle k \rangle=20$.
 Symbols correspond to averages over $100$ numerical simulations under the following conditions:
 \textbf{(a)} constant population and constant activity pdf $F(a)=\delta(a-a_0)$, \textbf{(b)} population growth and constant activity pdf $F(a)=\delta(a-a_0)$
 \textbf{(c)} constant population and power-law activity pdf $F(a)\propto a^{-1.5}$, \textbf{(d)} population growth and power-law activity pdf $F(a)\propto a^{-1.5}$.}
 \label{fig4}
\end{figure}

As we have said before, social networks are usually characterized by strong two-nodes degree correlations and strong transitivity
 (i.e., high probability that the friends of my friends are also my friends).  
An indirect measure of two-nodes degree correlation can be obtained from the neighbours average degree as a function of the
node degree $\bar{k}_{nn}(k)$, formally defined as \cite{Pastor-Satorras2001}:
\begin{equation}\label{eq13}
 \bar{k}_{nn}(k)=\sum_{k'}k'P(k'|k)
\end{equation}
where $P(k'|k)$ is the conditional probability that a node of degree $k$ is connected to another of degree $k'$. Thus, when $\bar{k}_{nn}(k)$
grows with $k$ we say the network has degree assortativity. \textbf{GSG} model yields networks with a clear degree assortativity 
under population growth regime, both for constant and power-law activity pdf $F(a)$, as shown in \fref{fig4}-b and \ref{fig4}-d.
In contrast, networks yielded by \textbf{GSG} model under constant population show a flat plot for $\bar{k}_{nn}(k)$, 
as would be expected in the case of Erd\"os-R\'enyi (ER) networks (random networks with Poisson degree distribution) for which $\bar{k}_{nn}(k')=\langle k\rangle$
 $\forall k'$ (see \fref{fig4}-a and \ref{fig4}-c). 

Another practical measure associated with transitivity can be given by the average clustering coefficient as a function of degree 
$\bar{C}(k)$, defined as:
\begin{equation}\label{eq14}
 \bar{C}(k)=\frac{1}{N_k}\sum_{i\in \mathrm{Deg}(k)}C_i,
\end{equation}
where $\mathrm{Deg}(k)$ is the set of all nodes of degree $k$, with $N_k$ its cardinal. In \eref{eq14} $C_i$ represent the 
local clustering for node $i$, defined as the fraction of edges between neighbours of node $i$ relative to its maximum number
$k_i(k_i-1)/2$ and reads
\begin{equation}
 C_i=\sum_{j,k \in N_{nn}(i)}\frac{a_{jk}}{k_i(k_i-1)},
\end{equation}
being $N_{nn}(i)$ the set of neighbours of node $i$ and $a_{jk}$ the elements of adjacency matrix $\mathbf{A}$, 
such that $a_{jk}=1$ ($a_{jk}=0$) if there is (not) an edge between nodes $i$ and $j$. Clustering $C_i$ is related with
the probability of triangles occurrence with node $i$ as one of its vertices.

Real social networks usually exhibit an scaling law for the average clustering as a function of degree $k$ 
\begin{equation}
 \bar{C}(k) \sim k^{-\beta}
\end{equation} 
where the observed exponents meet $\beta\lesssim 1$ \cite{Thurner13}.
\begin{figure}
 \centering
 \includegraphics[width=15 cm,height=13 cm,keepaspectratio=true]{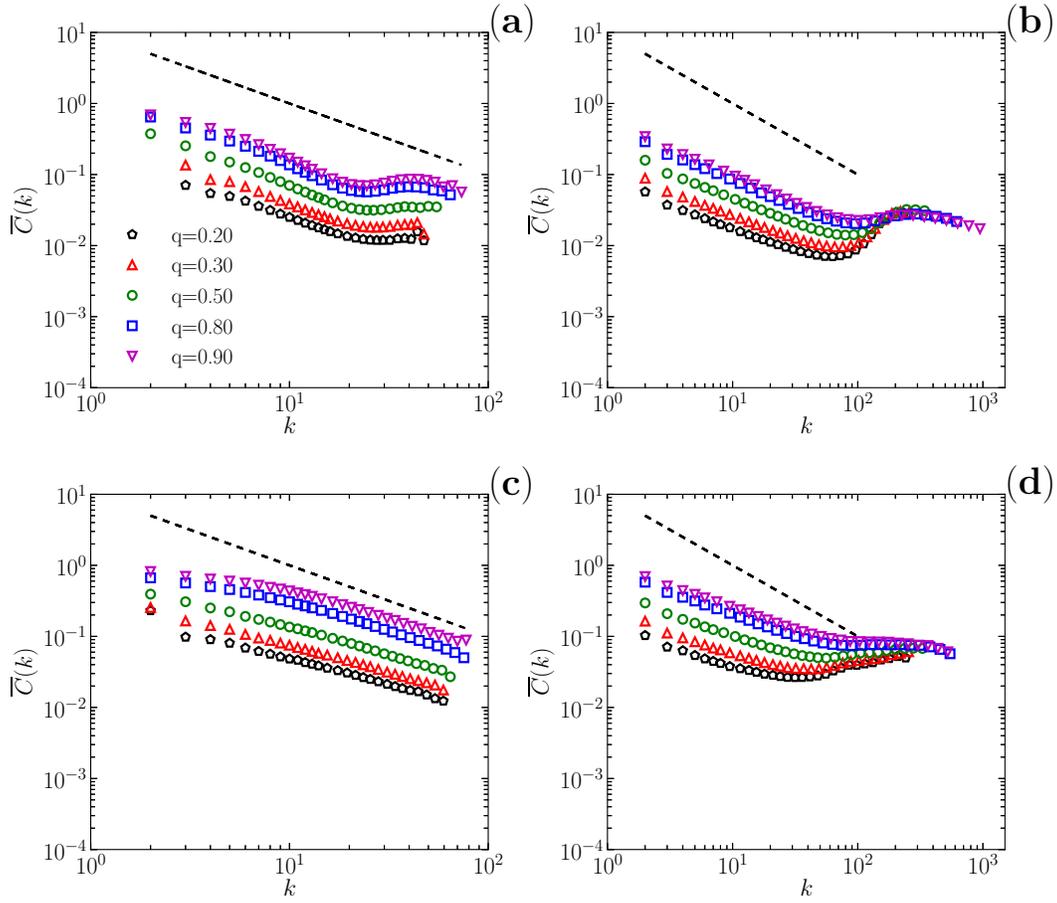}
 \caption{ Average local clustering $\bar{C}(k)$ as a function of degree $k$ for
 networks obtained by \textbf{GSG} model in all cases with final population $N=10^5$ and $\langle k \rangle=20$.
 Symbols correspond to averages over $100$ numerical simulations under the following conditions:
 \textbf{(a)} constant population and constant activity pdf $F(a)=\delta(a-a_0)$, \textbf{(b)} population growth and constant activity pdf $F(a)=\delta(a-a_0)$
 \textbf{(c)} constant population and power-law activity pdf $F(a)\propto a^{-1.5}$, \textbf{(d)} population growth and power-law activity pdf $F(a)\propto a^{-1.5}$.
 Dashed lines corresponding to the scaling-law $\bar{C}(k)\sim k^{-1}$ are plotted as reference.}
 \label{fig5}
\end{figure}
This scaling law behaviour is also captured by \textbf{GSG} model as shown in \fref{fig5}, where we have plotted
in all cases a dashed line with slope $-1$ in log-log scale as reference. This fact also shows that networks obtained by 
\textbf{GSG} model under constant population are far to be ER random networks as \fref{fig4}-a and \ref{fig4}-c might have suggested.
If that had been the case, $\bar{C}(k)$ would be independent of node degree $k$, as can be seen from its 
exact expression for random networks \cite{Dorogovtsev04}:
\begin{equation}
 \bar{C}(k)=\frac{(\langle k^2\rangle-\langle k\rangle)^2}{N\langle k\rangle^3}.
\end{equation}

\begin{figure}
 \centering
 \includegraphics[width=8 cm,height=10 cm,keepaspectratio=true]{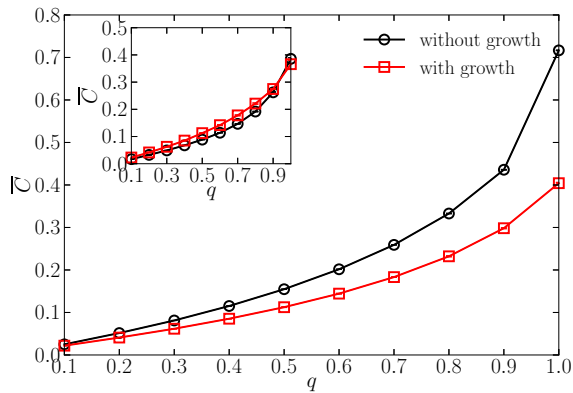}
 \caption{ Average clustering coefficient $\bar{C}$ in terms of \textbf{TC} probability.
 All results correspond to averages over $100$ realizations of \textbf{GSG} with $N=10^4$, $\langle k\rangle=20$ and $F(a)\propto a^{-1.5}$. 
 Inset: results for \textbf{GSG} model with the previous parameters but constant activity $F(a)=\delta(a-a_0)$.}
 \label{fig6}
\end{figure}

Finally, from the total average clustering coefficient definition:
\begin{equation}
 \bar{C}=\frac{1}{N}\sum_{i=1}^{N}C_i=\sum_{k}P(k)\bar{C}(k),
\end{equation}
we can confirm the expected growing nature of $\bar{C}$ with the \textbf{TC} probability $q$, as shown in \fref{fig6}.

\section{Cases of Study}\label{cases}
In order to illustrate the correspondence with \textbf{GSG} model, 
we analyze two experimental datasets reflecting human relationships networks in very different context: 
face-to-face encounters in a closed gathering and friendship relations in an online social network. 
For the first case, individual face-to-face contacts are detected with a time resolution of $20$ seconds and within a distance 
of $\sim 1$ meter, through wearable active radio-frequency identification devices (RFID)
placed on the chest of participants \cite{Cattuto10}. Here we analyze the publicly available dataset for ACM Hypertext 2009 (\textbf{HT})
conference held in Turin, Italy, with $N=113$ 
nodes and $L=2196$ unweighted edges\cite{Sociopatterns,Isella11} (we only consider time-aggregated face-to-face contacts between 
participants along the first meeting day). 
The second case corresponds to a Facebook subgraph (\textbf{FG}) comprising $N=63731$ users
from New Orleans (with larger connected component of size $N_{cc}=63392\approx 0.995\times N$), 
interconnected by $L=817090$ undirected edges \cite{comment_Facebook} representing friendship relations between users, 
as collected in \cite{Viswanath09}.
As in the previous case, this dataset also provides time-resolved information through the birth times of new edges.
In contrast to \textbf{HT}, \textbf{FG} exhibit population growth due to the introduction of new users, 
in a context without spatial constraints.

\begin{figure}
 \centering
 \includegraphics[width=14 cm,height=15 cm,keepaspectratio=true]{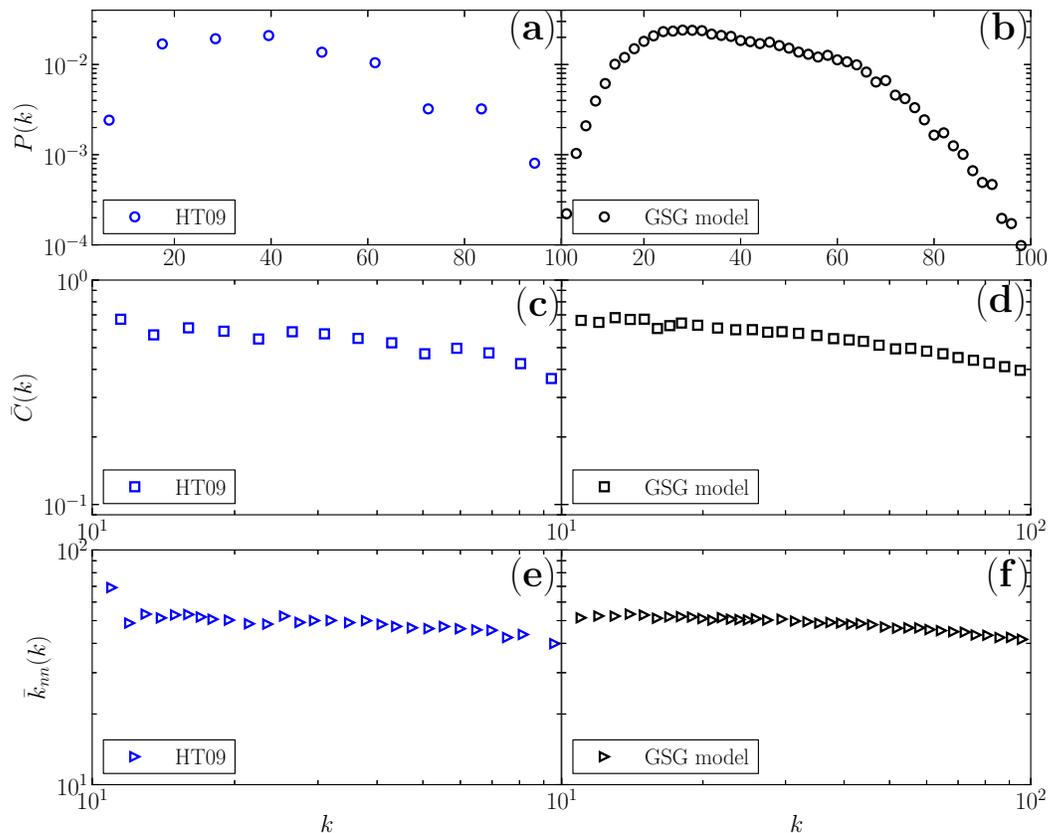}
 \caption{ Degree distribution $P(k)$, average clustering $\bar{C}(k)$ and neighbours average degree $\bar{k}_{nn}(k)$
  for \textbf{(a)},\textbf{(c)},\textbf{(e)} \textbf{HT} time-aggregated contacts network, and \textbf{(b)},\textbf{(d)},\textbf{(f)} \textbf{GSG} model under constant population 
  with $N=113$, $\langle k\rangle=38.9$, \textbf{TC} probability $q=0.8$ and power-law activity pdf with exponent $\xi=0.79$.}
 \label{fig7a}
\end{figure}

\begin{figure}
 \centering
 \includegraphics[width=14 cm,height=15 cm,keepaspectratio=true]{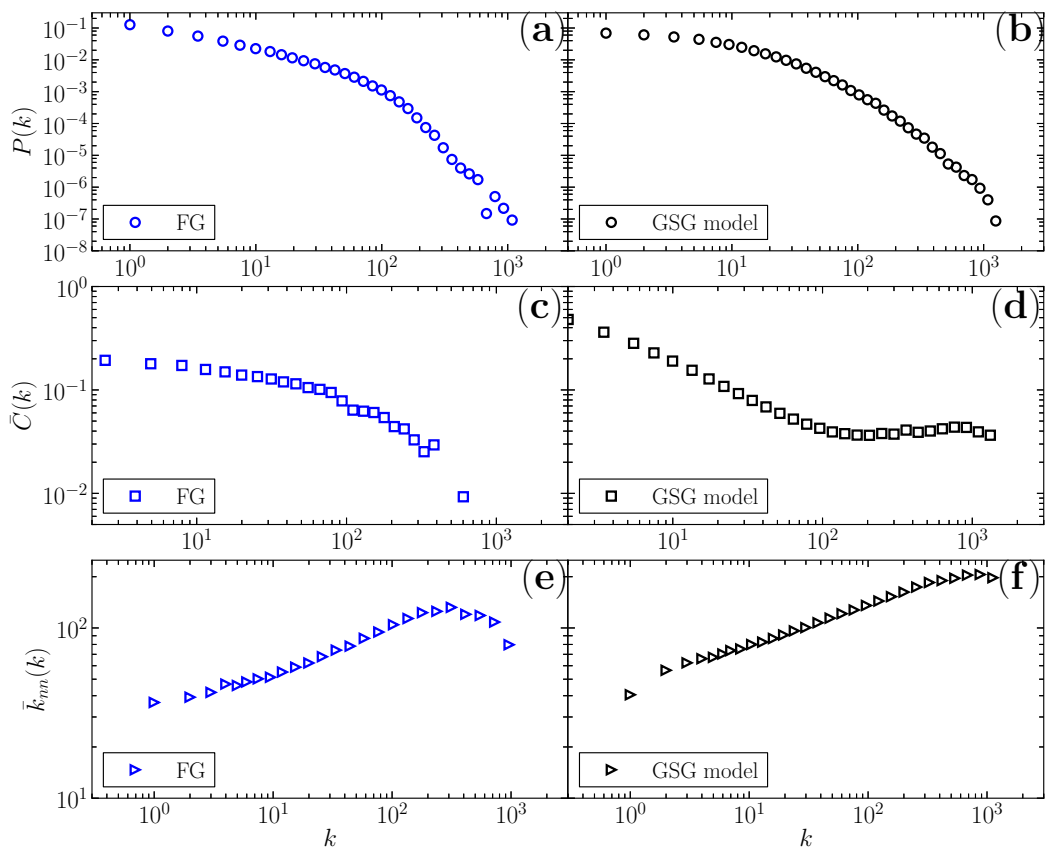}
 \caption{ Degree distribution $P(k)$, average clustering $\bar{C}(k)$ and neighbours average degree $\bar{k}_{nn}(k)$
  for \textbf{(a)},\textbf{(c)},\textbf{(e)} \textbf{FG} time-aggregated contacts network, and \textbf{(b)},\textbf{(d)},\textbf{(f)} \textbf{GSG} model under population growth
  with $N=63731$, $\langle k\rangle=25.6$, \textbf{TC} probability $q=0.8$ and power-law activity pdf with exponent $\xi=1.5$.}
 \label{fig7b}
\end{figure}

\subsection{Topological properties}
As can be seen in figures \ref{fig7a}-a, $P(k)$ for \textbf{HT} results narrow and short-tailed with 
small-$k$ Poissonian-like behaviour (see \fref{fig7a}-a) as obtained for \textbf{GSG} model under constant population 
with the same $N$ and $\langle k\rangle$ (see \fref{fig7a}-b).
Instead, \textbf{FG} network exhibit a heavy-tailed degree distribution with large-$k$ power-law behaviour 
$P(k)\sim k^{-\alpha}$ with $\alpha\approx 3.4$ as shown \fref{fig7b}-a, also compatible with a network generated by \textbf{GSG} model 
but now under population growth regime (see \fref{fig7b}-b). 
Additionally, the average clustering $\bar{C}(k)$ and the neighbours average degree $\bar{k}_{nn}(k)$ for \textbf{HT} and 
\textbf{FG} also exhibit the same qualitative behaviour of those generated by \textbf{GSG} model 
under constant and population growth regimes, respectively, as shown in \fref{fig7a}-(c-e) and \fref{fig7b}-(c-e). 

\subsection{Growth pattern: Triadic closure}\label{triadic}

Both \textbf{HT} and \textbf{FG} provide time-resolved information through the birth times of added edges. 
By virtue of this particular feature, we can address the problem of edges growth mechanism. 
Let $\mathbf{l_t}=(i,j)_t$ define an edge between nodes $i=\mathbf{l_t}(1)$ and $j=\mathbf{l_t}(2)$ emerged at time $t$, 
and let $d(\mathbf{l_t},t)$ the distance between nodes $\mathbf{l_t}(1)$ and $\mathbf{l_t}(2)$ at time $t$, 
immediately before the occurrence of $\mathbf{l_t}$. 
Those edges $\mathbf{l_t}$ for which $d_t(\mathbf{l_t},t)= 2$, called \textit{transitive edges}, are the product of a \textbf{TC} 
mechanism (or \textit{cyclic closure} for $d(\mathbf{l_t},2)> 2$ \cite{Watts06}). 

Then, we record $d(\mathbf{l_t},t)$ for each edge and define $N_d(T)$ as the aggregated number of edges $\mathbf{l_t}$ with 
$d(\mathbf{l_t},t)=d$ for $t<T$.
Let $d(\mathbf{l_t},t)=0$ when node $\mathbf{l_t}(1)$ or $\mathbf{l_t}(2)$ is a newcomer, and $d(\mathbf{l_t},t)=\infty$ when there is
not path between $\mathbf{l_t}(1)$ and $\mathbf{l_t}(2)$ previous to $\mathbf{l_t}$. 
Obviously $N_{d=1}(T)=0$ because multiple edges between nodes are forbidden. 

The distance distribution $P_T(d)$ is formally defined as
\begin{equation}
 P_T(d)=\lim_{L\rightarrow\infty}\frac{N_d(T)}{L} 
\end{equation}
with $L=\sum_{i\in\mathbb{N}_0} N_i(T)$ the total number of edges.
\begin{figure}
 \centering
 \includegraphics[width=15 cm,height=13 cm,keepaspectratio=true]{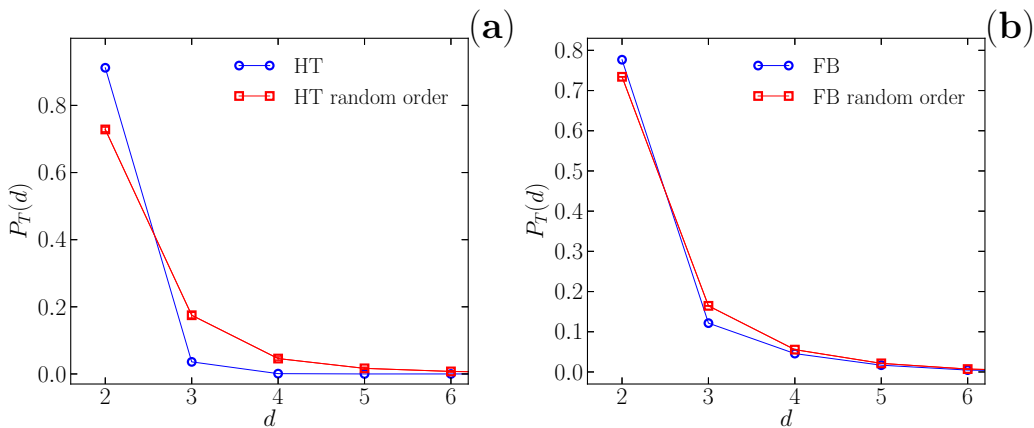}
 \caption{Distance probability distribution $P_T(d)$ as a function of distance $d$ (blue circles) and its average calculated 
 over $100$ random permutations of the actual time-ordered edges succession (red squares) for: \textbf{(a)} \textbf{HT} face-to-face dataset, 
 and \textbf{(b)} \textbf{FG} online friendship relations subgraph. Standard error bars are smaller than symbols.}
 \label{fig8}
\end{figure}
Clearly, $P_T(d)$ depends on the temporal ordering of $\lbrace\mathbf{l_t}\rbrace_{t\leq T}$ and this is why it provides 
valuable information about edges growth mechanism. 
In order to verify this statement, we compare $P_T(d)$ for the actual edges succession 
$\lbrace \mathbf{l_1}, \mathbf{l_2}, ..., \mathbf{l_T}\rbrace$, with the average 
$\langle P_T(d)\rangle_{rand}$ over $100$ random permutation  $\lbrace\mathbf{l}_{\sigma_t}\rbrace_{t=1,...,T}$, 
where $\sigma_{t}\in \mathrm{Perm}(T)$ is a random permutation function belonging to the group of all permutation of $T$ index 
$\mathrm{Perm}(T)$.
At this point is important to note that all succession $\{l_{\sigma_t}\}_{t=1,...,T}$ 
gives exactly the same final time-aggregated network at time $T$. 

Statistically significant differences are observed between $P_T(d)$ and $\langle P_T(d)\rangle_{rand}$ 
for both \textbf{HT} and \textbf{FG}, as shown in \fref{fig8}.
In particular, both plots in \fref{fig8} show $P_T(d)<\langle P_T(d)\rangle_{rand}$ for $d>2$ but $P_T(2)>\langle P_T(2)\rangle_{rand}$. 
The strong deviation for $d=2$ can be quantified through the $z$-score value, in this case defined as 
\begin{equation}
 z=\frac{P_T(2)-\langle P_T(2)\rangle_{rand}}{\sigma_{rand}},
\end{equation} 
resulting in $z_{HT}=32$ and $z_{FG}=185$. On the other hand, we have obtained a large fraction of transitive edges with 
$P_T(2)=0.91(2)$ for \textbf{HT} and $P_T(2)=0.77(4)$ for \textbf{FG}.
These facts states a clear memory effect in the mechanism of edges growth with a significant predominance of transitive edges.

\section{Summary and Discussion}\label{discussion}
In this work we have introduced an stochastic growth model (\textbf{GSG}) for social networks including an heterogeneous distribution of 
activity rate for the agents together with a \textbf{TC} mechanism for the attachment of new edges. 
The model allows to perform network growth processes by adding edges and, eventually, nodes depending on the population growth regime.
It also allows to work with different activity density functions within the same general framework. 
We show that \textbf{GSG} gives rise to time-aggregated networks having the main topological features expected 
for real social networks. 

We introduced a general analytical framework based on rate equation approach for node degree distribution under 
constant population and population growth regimes.
As has been previously reported \cite{Davidsen02,Holme2002,Vazquez03}, \textbf{TC} mechanism not only allows to increase 
the average clustering coefficient but also shapes the node degree distribution. In the particular case of \textbf{GSG} model, 
we shown that \textbf{TC} mechanism governs the large-$k$ behaviour of node degree distribution, giving place to 
exponential decay under constant population regime or power-law decay under population growth. On the other hand, low and mid-range $k$-values
are dominated by activity pdf together with the connectivity randomness.
Additionally, \textbf{GSG} model leads to time-aggregated networks with or without degree assortativity respectively 
for growing or constant population, and clustering with scaling-law behaviour
$C(k)\sim k^{-\beta}$ with $\beta\leq1$.

Recently, Karsai et al \cite{Perra14} have shown that the addition of long-term memory characteristic in agents affect their contacts
dynamic. 
In the same vein, the \textbf{GSG} model implicitly assume that each agent must remember all his previous ties in order to perform 
a \textbf{TC} step.
Thus, in this paper we have shown that agent's memory also leaves its mark on the time-aggregated network topology. 

The analysis of \textbf{HT} and \textbf{FG} time-resolved networks reinforce one of the main hypothesis behind \textbf{GSG} 
model, namely, that contacts between humans arise not totally at random but seem to favour the emergence of \textbf{TC} edges. 
Moreover, we have shown the qualitative agreement between \textbf{HT} and \textbf{FG} topological features with those 
obtained from the \textbf{GSG} model under constant population and population growth, respectively.
At this point we have to note the potential existence of some sort of bias on \textbf{FG} results induced by the Facebook's 
friend suggestions policy. However, this bias should not be present for the case of \textbf{HT}, what suggest the existence of
a possible behavioural mechanism exceeding any constraint imposed by an online platform.

Finally, our model open the way for further extensions in order to introduce a contacts dynamic 
obeying the topological structure of social networks.

\ack{We thanks to P. Balenzuela and M. Otero for helpful comments.
A.D.M. is grateful to Universidad de Buenos Aires for financial support through its postgraduate fellowship program. 
A.D.M and C.O.D acknowledge partial financial support from Universidad de Buenos Aires through its project UBACYT 2012-2015.}

\section*{References}
\bibliography{ref_Medus2014.bib}

\begin{thebibliography}{10}
\bibitem{comment_Facebook}
{Edges on Facebook can be considered as undirected because friendship requests
  must be explicitly accepted by the other party.}
\end{thebibliography}

\end{document}